\documentclass[10pt,a4]{article}  

\usepackage{graphicx}
\DeclareGraphicsExtensions{.eps}

\usepackage{mathptmx}   
\usepackage{latexsym}

\usepackage{amssymb} 
\usepackage{mathtools}

\renewcommand{\vec}[1]{{\bf #1}}

\newcommand{\N}[0]{\mathbb{N}}
\newcommand{\R}[0]{\mathbb{R}}
\newcommand{\Prob}[0]{\mathbb{P}}
\newcommand{\E}[0]{\mathbb{E}}
\newcommand{\ie}{i.e.\ }
\newcommand{\eg}{e.g.\ }
\newcommand{\vs}{vs.\ }
\newcommand{\cf}{cf.\ }
\newtheorem{proposition}{Proposition}
\newtheorem{theorem}{Theorem}
\newenvironment{Proof}{\textit{Proof}}{\hfill{}$\Box$}

\begin{document}
\title{Inter-Coder Agreement for Nominal Scales: A Model-based Approach}
\author{Dirk Schuster\thanks{Atos C-LAB, F\"urstenallee 11, D-33102 Paderborn,
    Germany. \hfill\hbox{}\hspace{\textwidth}\hbox{}\hspace{1.8em}Email: firstname.secondname@c-lab.de}}
\date{}
\maketitle

\begin{abstract}
  Inter-coder agreement measures, like Cohen's $\kappa$, correct the relative frequency of agreement
  between coders to account for agreement which simply occurs by chance.  However, in some
  situations these measures exhibit behavior which make their values difficult to interprete.  These
  properties, \eg the ``annotator bias'' or the ``problem of prevalence'', refer to a tendency of
  some of these measures to indicate counterintuitive high or low values of reliability depending
  on conditions which many researchers consider as unrelated to inter-coder reliability.  However,
  not all researchers agree with this view, and since there is no commonly accepted formal
  definition of inter-coder reliability, it is hard to decide whether this depends upon a different
  concept of reliability or simply upon flaws in the measuring algorithms.

  In this note we therefore take an axiomatic approach: we introduce a model for the rating of items
  by several coders according to a nominal scale.  Based upon this model we define inter-coder
  reliability as a probability to assign a category to an item with certainty.  We then discuss
  under which conditions this notion of inter-coder reliability is uniquely determined given
  typical experimental results, \ie relative frequencies of category
  assignments by different coders.
  
  In addition we provide an algorithm and conduct numerical simulations which exhibit the accuracy
  of this algorithm under different model parameter settings.

\end{abstract}

\section{Introduction}
\label{intro}
Measuring the agreement between the nominal ratings of a set of items by several coders or judges is
a common task in a number of disciplines like medical, psychological, and social sciences, content
analysis and marketing.  Simply measuring the percentage of agreement is not adequate as it does not
take into account agreement which simply occurs by chance.  There have been proposed a number of
inter-coder reliability measures to cope with this effect, the most prominent being $\kappa$
\cite{Cohen1960}, $\pi$ (\cite{Scott1955}, \cite{Fleiss1971}), $\alpha$ \cite{Krippendorff1980}, and
$S$ \cite{Bennett1954}, see \cite{Artstein2008} for a survey.

These measures are defined as ratios of chance-corrected numbers of observed agreement \vs maximal
agreement and differ in the way the chance-correction is taken into account.  The ways these
corrections are computed, give rise to some criticism of these measures, because they ``favor'' or
``penalize'' certain coder behaviors which are considered as inappropriate by some researchers.

Though usually not explicitly stated (\cf also \cite[p. 294]{Aickin1990}), the basic assumption is
that a coder either assigns a category by certainty resp.\ ``expert judgment'' (Brennan and
Prediger, \cite[p. 689]{Brennan1981}) or assigns some category without being absolutely sure about
his or her choice.  Obviously, it is not possible for an individual assignment to identify whether
the assignment was done by certainty or not, sometimes not even the rater himself or herself may be
sure about what the exact reasons for his or her choice are.

At one extreme point is the $S$-value which assumes a uniform distribution of categories when
``chance assignments'' occurs.  Scott's $\pi$ and Cohen's $\kappa$ on the other hand use ``marginal
distributions'', \ie the overall distribution of category assignments by each rater, to correct for
chance agreement.  Using marginal distributions may lead to incorrect chance correction since these
distributions also include assignments made by certainty and thus may also be more than marginally
influenced by the distribution of categories according to the population of items.  Using uniform
distribution on the other hand may underestimate chance agreement if there are categories that
coders hardly ever choose.  There exists a considerable literature on this subject, see \eg
\cite{Artstein2008}, \cite{Brennan1981}, \cite{DiEugenio2004}, \cite{Feinstein1990},
\cite{Gwet2002},\cite{Hsu2003} \cite{Kraemer1979}.

Obviously it is hard to reach at a consensus about which strategy a coder will follow in general
when category assignment is not done by certainty.  In our model we thus will not presume a certain
distribution to account for chance agreement.

Cohen's $\kappa$ exhibits a feature, usually called ``annotator bias'' which describes the fact that
$\kappa$ yields higher values when coders produce widely diverging marginal distributions than when
the marginal distributions are similar. See \cite[section 3.1]{Artstein2008} who support this
feature, \cite{DiEugenio2004}, \cite{Feinstein1990},\cite{Zwick1988} for criticism,
\cite{Warrens2010a} for a formal proof.  Scott's $\pi$, in contrast, uses the common marginal
distribution of the coders and so ``favors'' coders that produce similar marginal distributions.

In order to measure inter-coder reliability (in contrast to intra-coder, \ie test-retest
reliability) it is necessary that the experiment can be reproduced when conducted in the same way
with another group of coders (which of course may be restricted to a certain base population \eg
trained in some way, but not delimited to some particular individuals).  So an inter-coder
reliability measure should (approximately) yield the same value for every sufficiently large subset
of coders from the prescribed population of coders and the coders' marginal distributions may vary
according to some distribution which depends on the population of coders. 

Another debated fact is the prevalence problem, referring to the fact that some of these measures
($\kappa$,$\pi$,$\alpha$) produce low scores when one category is predominant among the ratings
(see \cite{Artstein2008}, \cite{DiEugenio2004}, \cite{Feinstein1990}, \cite{Gwet2002} for examples
and discussion).

There is some debate on this issue.  While \cite{DiEugenio2004}, \cite{Feinstein1990} and
\cite{Gwet2002} consider this as a weakness, it is justified by Artstein and Poesio with the
argument that ``reliability in such cases is the ability to agree on rare categories'' \cite[section
3.2]{Artstein2008}.  This latter argument is somewhat problematic for statistical measures which
usually are designed to exhibit typical not exceptional behavior.  In our model we will take an
approach which defines reliability as a property common to the category assignments and independent
of the relative frequency of the (``true'' or ``correct'') items' categories.  However it will turn
out that reliability can only be determined if not all items belong to one category.

The approach we take differs from these measures as we start with an axiomatic model-based
definition of inter-coder reliability, which will be a probability of some event.  This has the nice
side effect that the value of the reliability parameter can be stated as a probability of an idealized
coder's behavior and thus has a direct interpretation.

In addition basing the definition of inter-coder reliability upon such a model one may simulate coder
ratings with a known reliability parameter and thus may evaluate the accuracy of algorithms under
different setups.  We will do this in Section~\ref{sec:numerical-simulation} for the algorithm we
provide.

Though the author believes that the model used here is fairly general, there might be situations in
which it could be deemed unfeasible.  Here the explicit statement of the model's assumptions helps
to determine whether the model is acceptable in an experiment or not.  We will take a closer look at
some of the assumptions of the model and their possible impact on reliability results at the end of
the next section.

\section{The Model}
\label{sec:model}
We denote by $C = \{c_1,\ldots, c_m\}$ the (finite) set of $m$ categories, into which $N$ items, $\N_N$, are to be
classified by the $R$ raters, $\N_R$.  We use $\N_n$ to denote the natural numbers $\{1,\ldots,n\}$. 

The common assumptions for inter-rater agreement are (rephrased from \cite{Cohen1960}):
\begin{enumerate}
\item \label{enum:model-assumption-independent-items}The items are independent
\item \label{enum:model-assumption-category} The categories are independent, mutually exclusive, and exhaustive.
\item \label{enum:model-assumption-independent-raters} The raters operate independently
\end{enumerate}

Assumption~\ref{enum:model-assumption-category} that categories are exhaustive and mutually
exclusive implies that for every item there is one and only one ``correct'' category.  In other
words, assumption~\ref{enum:model-assumption-category} above implies the existence of a (usually
unknown) function $$\gamma\,:\,\N_N \rightarrow C.$$ We will sometimes call $\gamma(k)$ the ``true''
category associated with item $k$, without any philosophical implication of the term ``true''.

For each $c\in C$ let $N_c := \# \gamma^{-1}(c)$ denote the number of items whose true category is
$c$, and write $\tau_c := \frac{N_c}{N}$ for the relative frequency of these items.

If a coder rates an item he or she may either be sure about the category to be chosen or not.  If
the coder is sure about the item's category it seems natural to assume that the coder will assign
this category to the item (so we assume that the coders will not cheat but will assign a category to the
best of their knowledge).

Now, what happens in the case the coder is not completely sure about the category to assign?  In
this case, considering a large set of such items, we will observe a certain relative frequency for
the categories to be chosen.  In general it is hard to know which strategy the coder will take and
this is frequently debated in the context of Cohen's $\kappa$.  Coders might follow some ``base
rate'' \ie are guided by some assumption about the distribution of categories in the population of
items, or may choose the category according to a uniform distribution on the set of categories (\cf \eg
\cite{Bennett1954}, \cite{Hsu2003}).

There are certainly good reasons for many of these assumptions and it is probably also dependent
upon the field of research (\eg medical diagnosis \vs speech analysis), upon the kinds of items, the
professional background and education of the raters (\eg scholars \vs laymen) and many more properties.
Hence we will not assume any particular distribution but only assume that such a distribution
exists.

To formalize we thus assume that given an item $k$ a rater recognizes the true category $\gamma(k)$
with a probability $\beta$.  If the coder fails to recognize it he or she assigns a ``random''
category with some unknown distribution. The assumption
\ref{enum:model-assumption-independent-items} above suggests to model these actions by independent
random variables.

So formally let $Z_{k}$ be 0-1-valued, $Y_{k}$ be $C$-valued independent random variables
$k\in\N_N$ and assume that both families are identically distributed.

We define a coder's rating of item $k\in \N_N$ by the outcome of the random variable $X_k$ given by
\begin{equation}
  \label{eq:rater-model}
  X_k := 
  \begin{cases}
     \gamma(k) &, \text{ if } Z_{k} = 0\\
     Y_{k}  &, \text{ if } Z_{k} = 1.
    \end{cases} 
\end{equation}

We let $\beta := \Prob(Z_{k} = 0)$ and $p_c := \Prob(Y_{k} = c)$.

It is immediate from the definition that 
\begin{equation}
  \Prob(X_{k} = c) = \beta\delta_{c,\gamma(k)} + (1-\beta) p_c\label{eq:basic-prob-Xk=c},
\end{equation}
so the distribution of $X_{i,k}$ is a mixture of the atomic distribution at $\gamma(k)$ and
$\vec{p} = (\Prob^{Y_{k}}(c))_{c\in C}$ with mixture parameter $\beta$. (Here $\delta$ is
Kronecker's delta, \ie $\delta_{x,y} = 1$ if $x=y$ and $0$ otherwise.)

For convenience let us call this model the \textit{coder model with parameters $(\beta, \gamma,
  \vec{p})$}, where $\vec{p} = (p_c)_{c\in C}$.  Throughout this note we will tacitly let $\N_N$
denote the domain and $C$ the codomain of $\gamma$.  A family of independent $C$-valued random
variables $(X_k)_{k\in \N_N}$, which satisfies \eqref{eq:basic-prob-Xk=c} is called a \textit{coder
  process} for the coder model.

According to the assumption \ref{enum:model-assumption-independent-raters} above several coders are
modeled by independent families $(X_{i,k})_{k\in\N_N}$, where the subscript $i$ refers to the
coder.

If a rater chooses to assign category $c$ to an item $k$ he or she may either be certain about the
items category or may be uncertain and assigns $c$ by chance only.  Gwet \cite[section 4]{Gwet2008}
uses this same interpretation of the rating process.  In our model certainty occurs when the coder
chooses the category according to $\gamma(k)$, \ie when $Z_k=0$.  So it seems reasonable to use the
probability $\beta = \Prob(Z_k=0)$ as agreement indicator, let us call it the \textit{reliability
  parameter} of the coder model.  Of course an assignment to category $\gamma(k)$ also occurs when
$Y_k = \gamma(k)$, which happens with probability $p_{\gamma(k)}$.

Aickin \cite{Aickin1990} also used a mixture model to study inter-coder reliability.  In our
notation the mixture distribution in Aickin's model is the distribution
$$(c_1,c_2) \mapsto \E(\frac{1}{N}\#\{ i\in \N_N : X_{i,1} = c_1 \} \frac{1}{N}\#\{ i\in \N_N : X_{i,2} = c_2 \})$$
which is not a mixture distribution in our model, \cf \eqref{eq:def-e_2c1c2}, so our model is
different from Aickin's.

There are three features of this model which may need a second look:

The first feature is that coders are modeled by identically distributed families of random
variables.  This might seem oversimplifying since generally every coder may have his or her own
preference.  Actually this feature touches the controversy about ``annotator bias''. 

As we already discussed in the introduction, inter-coder reliability in contrast to intra-coder
reliability is only present if the experiment can be reproduced with different coders from some
coder population.  Our model uses the parameters $\beta$ and $\vec{p}$ to characterize the coder
population.

The second feature that may deserve closer consideration, is concerned with the \textit{a priori}
distribution $\vec{p}$ being independent of the item in question.  Actually often one may arrive at
the situation where for a particular item the coders easily may rule out some categories but are
doubtful about some others.  In this situation assuming that the \textit{a priori} distribution is
the same for all items is indeed oversimplifying.  Without this assumption, however, the model would
be completely useless.  Indeed, if $\vec{p}$ would be dependent on the item $k$ we could simply put
$\vec{p}^{(k)}$ to the distribution of categories obtained for this item and find out that every
outcome could be obtained with reliability parameter $\beta=0$, \ie by pure randomness.

If in some experimental setup the independence of $\vec{p}$ on the item would be deemed a relevant
issue, it would be advisable to split the set of items into subsets such that the \textit{a priori}
distribution could be considered the same for all items in each of the subsets.  

The third feature which deserves attention is that the probability to identify item $k$ as belonging
to category $\gamma(k)$ is independent of $\gamma(k)$, \ie that $\beta$ is considered independent of
$c$.  It is easy to imagine a situation where some subset of categories are more easily
distinguished from each other than for another subset.  In this situation it would indeed be more
appropriate to assume $\beta$ to be dependent of $c$.  However this would entail the necessity to
report several values as reliability parameter, which would make comparisons more difficult.

Even here one should cope with this feature by a careful design of the experiment (choice of
categories).  We will return to this aspect later (following
Proposition~\ref{prop:category-mapping}).

\section{Inter-Coder Agreement}
\label{sec:intercoder-agreement}

According to our model inter-coder reliability is the parameter $\beta$ in
\eqref{eq:basic-prob-Xk=c} which, since $\gamma$ is unknown, is not directly observable in
experiments.  In experiments only relative frequencies of category assignments can be observed, \ie
we can observe $\frac{1}{N}\#\{ i\in \N_N : X_{i,1} = c_1,\ldots{},X_{i,r} = c_r\}$ or, idealized,
the expectation values of it.  In the present section we will discuss under which conditions $\beta$
can be uniquely determined from these expectation values.

Throughout this section we will frequently use the following relations, the proof of which is
obvious from \eqref{eq:basic-prob-Xk=c} and the independence of $X_{i,k}$.

\begin{align}
  e_{1,c} &:= \E\left( \frac{1}{N}\#\{ k\in\N_N: X_{k} = c\} \right) = \beta\tau + (1-\beta)p_c\label{eq:def-e_1c}\\
  e_{2,c_1,c_2} &:= \E\left( \frac{1}{N}\#\{ k\in\N_N: X_{1,k} = c_1, X_{2,k} =
    c_2\}\right)\nonumber\\
&= \beta^2\delta_{c_i,c_2}\tau_{c_1} + \beta(1-\beta) ( \tau_{c_1}p_{c_2} + \tau_{c_2}p_{c_1} ) + (1-\beta)^2 p_{c_1} p_{c_2}\label{eq:def-e_2c1c2}\\
  e_{2,c} &:= e_{2,c,c} = \beta^2\tau_c +2 \beta (1-\beta) \tau_c p_c + (1-\beta)^2 p_c^2\label{eq:def-e_2c}\\
  e_{3,c} &:= \E\left( \frac{1}{N}\#\{ k\in\N_N: X_{1,k} = X_{2,k} = X_{3,k} =
    c\}\right)\nonumber \\
    &= \beta^3\tau_c + 3\beta^2(1-\beta)\tau_c p_c  + 3\beta(1-\beta)^2\tau_c p_c^2 + (1-\beta)^3p_c^3\label{eq:def-e_3c}
\end{align}

Our first result shows that it is not always possible to identify $\beta$ from the coder's ratings.

\begin{proposition}
  \label{proposition:single-category-case}
  Let $(\beta,\gamma, \vec{p})$ be a coder model and assume that there is $c_0\in C$ such that $\gamma(k) = c_0$ for
  all $k\in \N_N$.  Then for every $\beta' \leq \beta + (1-\beta)p_{c_0}$ there is a coder model $(\beta', \gamma,
  \vec{p'})$ such that
  \begin{equation}\label{eq:proposition:single-category-case:prob-equality}
        \beta\delta_{c,\gamma(k)} + (1-\beta)p_c = \beta' \delta_{c,\gamma(k)} + (1-\beta') p_c'     
  \end{equation}
  for all $c\in C$, $k\in\N_N$.
\end{proposition}

\begin{Proof}
  Given $\beta' \leq \beta + (1-\beta)p_{c_0}$, we only have to show the existence of a vector $\vec{p'}
  \in [0,1]^m$ with $\sum_{c\in C} p_c = 1$ such that \eqref{eq:proposition:single-category-case:prob-equality} holds.

  Assume first that $\beta'=1$. Then $1=\beta' \leq \beta+ (1-\beta)p_{c_0} \leq 1$ so 
  \begin{equation}
    \label{eq:proposition:single-category-case:1}
    (1-\beta)(p_{c_0} - 1) = 0
  \end{equation}
  Hence either $\beta=1$ or $p_{c_0} = 1$. In the first case the statement is trivially satisfied and in the
  second case we may set $p_c' = p_c$ for all $c\in C$ and obtain either 1 (if $c=c_0$) or 0 on both sides of
  \eqref{eq:proposition:single-category-case:prob-equality}, proving the statement in this case.
  
  Now assume $\beta' < 1$. Then
  \begin{align*}
    \beta' & \leq \beta + (1-\beta) p_{c_0} = \beta + (1-\beta) (1 - \sum_{c\not=c_0} p_c) \leq 1 - (1-\beta)  p_c
  \end{align*}
  for all $c\in C\setminus \{c_0\}$. Hence  $(1-\beta) p_c \leq 1-\beta'$, so defining
  $$p_c' := \frac{1-\beta}{1-\beta'}p_c$$
  we obtain $p_c' \in [0,1]$, for $c\not= c_0$. Also define
  $$p_{c_0}' := \frac{\beta-\beta'+ (1-\beta)p_c}{1-\beta'}.$$
  Then $p_{c_0}' \geq 0$ by the condition on $\beta'$ and 
  $\beta-\beta' + (1-\beta)p_c \leq \beta - \beta' + (1-\beta) = 1 - \beta'$
  shows $p_{c_0}' \leq 1$. Finally, 
  $$\sum_{c\in C} p_c' = p_{c_0}' + \sum_{c\not=c_0} p_c' = \frac{\beta-\beta'}{1-\beta'} +  \frac{1-\beta}{1-\beta'}p_{c_0} + \frac{1-\beta}{1-\beta'}(1-p_{c_0}) = 1$$
  completes the proof.
\end{Proof}

Note that in Proposition~\ref{proposition:single-category-case} we may always choose $\beta'=0$, so the rating cannot
be distinguished from a completely random one, but of course at the cost of a distribution $\vec{p'}$ possibly far
from uniform. In the case $\beta=1$, $\beta'=0$ the distribution $\vec{p'}$ is atomic at $c_0$, which somewhat
challenges the intuition of ``random agreement''.

On the other hand, unless we \textit{know} that $\#\gamma(\N_N) > 1$, we are actually unable to
determine the reliability parameter $\beta$.

\begin{proposition}
  \label{prop:case-tau-known} Let $(\beta,\gamma, \vec{p})$ be a coder model and assume that
  $\tau_{c} < 1$ for all $c\in C$. Then the following holds
  \begin{enumerate}
  \item\label{prop:case-tau-known:item-tau-or-beta=0} if $e_{2,c_0} = e_{1,c_0}^2$ for some
    $c_0$ then either $\tau_{c_0} = 0$ or $\beta=0$.
  \item\label{prop:case-tau-known:item-beta=0}  $e_{2,c} = e_{1,c}^2$ for all $c\in C$ if and only if $\beta=0$
  \item\label{prop:case-tau-known:item-beta-soln}  if $e_{2,c_0} \not= e_{1,c_0}^2$ for some $c_0$
    then $e_{2,c_0} > e_{1,c_0}^2$ and
    \begin{equation}
      \label{eq:prop:case-tau-known:beta-soln}
      \beta = \sqrt{\frac{e_{2,{c_0}} - e_{1,{c_0}}^2}{\tau_{c_0} (1-\tau_{c_0})}}
    \end{equation}
    and $p_{c_0}$ is given by
    \begin{equation}     
      p_{c_0} = \frac{e_{1,{c_0}} - \beta\tau_{c_0}}{1-\beta}\label{eq:prop:case-tau-known:p_c0-soln}
    \end{equation}
 \end{enumerate}
\end{proposition}

\begin{Proof}
  From \eqref{eq:def-e_1c} and \eqref{eq:def-e_2c} for any $c\in C$ 
  $$e_{1,c}^2 = \beta^2\tau_{c} + 2\beta(1-\beta)\tau_{c} p_{c} + (1-\beta)^2 p_{c}^2$$ 
  so
  \begin{equation}
    \label{eq:proof-uniform-category-case-joint-agreement-excess}
    e_{2,c} - e_{1,c}^2 = \beta^2\tau_{c} (1-\tau_{c}) 
  \end{equation}
  Since by assumption $\tau_c<1$ equation
  \eqref{eq:proof-uniform-category-case-joint-agreement-excess} shows
  part~\ref{prop:case-tau-known:item-tau-or-beta=0}. And since $\sum_c \tau_c = 1$
  there is $0 < \tau_{c_0} < 1$ for some $c_0 \in C$ proving
  part~\ref{prop:case-tau-known:item-beta=0}.  Solving
  \eqref{eq:proof-uniform-category-case-joint-agreement-excess} for $\beta$ and \eqref{eq:def-e_1c}
  for $p_{c_0}$ completes the proof.
\end{Proof}

One application of this result is, that one may determine the range of the distribution $p_c$ from
the results of a pre-study with a carefully chosen set of items with known ``true'' categories which
meet the assumption $0 < \tau_c < 1$ for all $c\in C$. Once we know the range, \ie $min_{c\in
  C}(p_c)$ and $max_{c\in C}(p_c)$ of the \textit{a priori} distribution and can reasonably assume that
it does not change for an arbitrary set of items, we can estimate the reliability parameter for
arbitrary distribution $\vec{\tau}$ of true categories using the following

\begin{proposition}
  \label{proposition:beta-estimate} Let $(\beta,\gamma, \vec{p})$ be a coder model, assume that $\pi_0 \leq p_c
  \leq \pi_1 < 1$ for $c\in C$ and define $e_2 := \sum_{c\in C} e_{2,c}$. Then the following estimate holds
  $$\sqrt{\frac{max(0,e_2 - \pi_1)}{1-\pi_1}} \leq \beta \leq \sqrt{\frac{e_2 - \pi_0}{1-\pi_0}}$$
\end{proposition}

\begin{Proof}
    From \eqref{eq:def-e_2c} we see that 
    \begin{align}
    e_2 &=  \sum_{c\in C}
             \left(
                   \beta^2\tau_c + 2\beta(1-\beta) \tau_c p_c  + (1-\beta)^2 p_c^2
             \right)\\
        &= \beta^2 + 2\beta(1-\beta) \sum_{c\in C} \tau_c p_c  + (1-\beta)^2 \sum_{c\in C} p_c^2
    \end{align}
    Now, since $0 \leq \tau_c$ we may estimate $\pi_0 \tau_c \leq  \;p_c \tau_c \leq \pi_1\tau_c$ and $\pi_0 p_c \leq  \;p_c^2 \leq \pi_1 p_c$
    and since $\sum_{c\in C} \tau_c = 1 = \sum_{c\in C} p_c$ thus obtain 
    \begin{align}
      \pi_0  = \sum_{c\in C} \pi_0 \tau_c  \leq &\sum_{c\in C}\tau_c p_c  \leq \pi_1, \text{ and }\\
      \pi_0  \leq &\sum_{c\in C} p_c^2 \leq \pi_1
    \end{align}
    Thus we may $e_2$ estimate from below
    \begin{align}
      e_2 \geq \beta^2 + 2\beta(1-\beta)\pi_0  + (1-\beta)^2 \pi_o = (1-\pi_0) \beta^2 + \pi_0
    \end{align}
    (which implies $e_2 - \pi_0 > 0$) and similarly from above (replacing $\pi_o$ by $\pi_1$).
    Since by assumption $1-\pi_1 > 0$ and  $1-\pi_0 > 0$ we obtain the desired
    estimates.
\end{Proof}

Observe, that the preceding proposition does not assume anything about $\tau_c$, it even holds if
$\tau_{c_0} = 1$ for some $c_0$. 

If $p_c$ is the uniform distribution we may put $\pi_0 = \pi_1 = \frac{1}{m}$ in the preceding result and
obtain the equality $\beta^2 = \frac{e_2 - \frac{1}{m}}{1 - \frac{1}{m}}$ which is the $S$-value of Bennett,
Alpert and Goldstein \cite{Bennett1954}.

The $S$-value has been criticized by Scott \cite{Scott1955} that it could be increased by adding
spurious categories which would never or hardly ever be used.  But, as Scott also notes, such a
modification would contradict the assumption of uniform distribution for $p_c$, hence by such a
modification $\beta$ can no longer be determined by the $S$-value formula.  We may however use the
preceding proposition to obtain estimates for $\beta$: if one adds a category $c_0$ that a coder
wouldn't use, the \textit{a priori} probability $p_{c_0}$ is $0$ and so is the minimum, hence we
would obtain the inequality
$$\sqrt{\frac{e_2 - \pi_1}{1-\pi_1}} \leq \beta \leq \sqrt{e_2}.$$
Since $\sqrt{\frac{e_2 - \pi_0}{1-\pi_0}} \leq \sqrt{e_2}$, adding such a spurious category results
in a larger possible interval for $\beta$, \ie a worse estimate. 

If we even know the distribution $\vec{p}$ we are able to compute $\beta$ exactly.  The same is true
if the \textit{a priori} probability $p_c$ matches the item category distribution $\tau_c$.  This is
the content of the following

\begin{proposition}
  \label{proposition:special-cases} Let $(\beta,\gamma, \vec{p})$ be a coder model and assume that $\tau_c
  < 1$ for all $c\in C$.
  \begin{enumerate}\item\label{proposition:special-cases:p-known}  If $\vec{p}$ is known, $\beta$ can be computed as
    \begin{equation}\label{eq:beta-for-p-given}
      \beta =
      \begin{cases}
        0\text{, if }p_c = 1 \text{ and } e_{1,c}=1 \\
        \frac{1- 2e_{1,c}+e_{2,c}}{1-e_{1,c}}\text{, if }p_c = 1 \text{ and } e_{1,c}<1\\
        1 - \frac{1}{2} \left( \frac{1-e_{1,c}}{1-p_c} + \frac{e_{1,c}}{p_c} \right) +
        \sqrt{\frac{1}{4} \left( \frac{1-e_{1,c}}{1-p_c} + \frac{e_{1,c}}{p_c} \right)^2 -
          \frac{e_{1,c}-e_{2,c}}{p_c (1-p_c)}}\text{, if }0 < p_c < 1
      \end{cases}
    \end{equation}
     \item\label{proposition:special-cases:p-equals-tau} If $0 < \tau_c = p_c < 1$ for some $c\in C$.  Then
    \begin{equation}
      \label{eq:beta-for-tau-eq-p}
      \beta = \sqrt{\frac{e_{2,c}-e_{1,c}^2}{e_{1,c}(1-e_{1,c})}}
    \end{equation}
  \end{enumerate}
\end{proposition}

\begin{Proof}
  From \eqref{eq:def-e_2c} and \eqref{eq:def-e_1c} we obtain
  \begin{align}
    e_{2,c} &= (\beta + 2 (1-\beta)p_c) (e_{1,c}- (1-\beta)p_c) + (1-\beta)^2p_c^2 \nonumber\\
            &= (\beta-1)^2 p_c (1-p_c) + (\beta-1) (e_{1,c} +p_c - 2 p_c e_{1,c}) + e_{1,c} \nonumber
  \end{align}
  hence
  \begin{equation}
   f(\beta) := (\beta-1)^2 p_c (1-p_c) + (\beta-1) \left( p_c (1-e_{1,c}) + e_{1,c} (1-p_c)\right) + (e_{1,c}-e_{2,c}) = 0\label{eq:eqn-for-beta-if-p-given}
  \end{equation}  
  First observe, that $e_{1,c} - e_{2,c} = \sum_{c'\in C} e_{2,c,c'} \geq e_{2,c,c} - e_{2,c} = 0$
  Since $\sum_{c\in C} p_c = 1$ there is $c\in C$ with $p_c > 0$. 

  If $p_c=1$ then $f$ is linear. The linear term also vanishes, if in addition $e_{1,c}=1$.  In this
  case $0 = e_{1,c}-1 = \beta(\tau_c - 1)$, so $\beta=0$.  On the other hand, if $e_{1,c} \not= 1$
  we can solve \eqref{eq:eqn-for-beta-if-p-given} for $\beta$ and obtain the second case of
  \eqref{eq:beta-for-p-given}.

  Now assume $0 < p_c < 1$ then $e_{2,c}-e_{1,c}^2 \geq 0$ by
  Proposition~\ref{prop:case-tau-known}\ref{prop:case-tau-known:item-beta-soln} and
  \begin{align*}
    f(0) &= - (p_c - e_{1,c})^2 - (e_{2,c}-e_{1,c}^2) \leq 0\\
    f(1) &= e_{1,c}-e_{2,c} \geq 0
  \end{align*}
  so there is one zero of $f$ in the interval $[0,1]$ and one in $]-\infty,0]$.  Solving
  \eqref{eq:eqn-for-beta-if-p-given} for $\beta$ and discarding the lower solution yields
  \eqref{eq:beta-for-p-given}.

  Finally, if $\tau_c = p_c$ we have $e_{1,c} = p_c\in\; ]0,1[$, hence $e_{1,c} (1-e_{1,c}) \not= 0$
  and
  $$e_{2,c} - e_{1,c}^2 = \beta^2 p_c(1-p_c) = \beta^2 e_{1,c} (1-e_{1,c})$$
  immediately shows \eqref{eq:beta-for-tau-eq-p}.
\end{Proof}

Assume that the population of items is a representative sample from the universe of items and that
the coders know about the distribution of categories (``base rate'') in the universe (such a situation seems not
uncommon in medical or psychological diagnostics) then
part~\ref{proposition:special-cases:p-equals-tau} provides a simple method to compute reliability.
If the coder's assumption on the base rate differs from the ``true'' category distribution $\beta$
can be computed from part~\ref{proposition:special-cases:p-known}.  

As was announced in the introduction our model does not share the ``annotator bias'' property, which
is obvious from the definition of the model.  It is also known that $\kappa$ may be increased or
decreased by combining categories (see \cite{Warrens2012}).  Therefore it is worth recording the
following proposition which shows that $\beta$ does not change when combining categories or adding
spurious ones.

\begin{proposition}\label{prop:category-mapping}
  Let $(\beta,\gamma, \vec{p})$ be a coder model with coder process $X_k$. Let $C'$ be a finite set
  and let $\Phi\,:\,C \rightarrow C'$ be some map. Let $X_k' := \Phi\circ X_k$, $\gamma\,' =
  \Phi\circ \gamma$ and for every $c'\in C'$ let $p_{c'}' = \sum_{c\in \Phi^{-1}(c')} p_c$ (with the
  understanding that $p_{c'}' = 0$ whenever $\Phi^{-1}(c') = \emptyset$) .

  Then $X_k'$ is a coder process for $(\beta, \gamma\,',\vec{p'})$, \ie 
  \begin{equation}
        \Prob(X_k' = c') = \beta \delta_{c',\gamma'(k)} + (1-\beta)p_{c'}'\label{eq:prop:category-mapping} 
  \end{equation} 
  for all $k\in N_N$, $c'\in C'$.
\end{proposition}

Note that the definition of $p_{c'}'$ in the proposition just defines the distribution of
\hbox{$\Phi\circ Y_k$} on $C'$ with $Y_k$ from \eqref{eq:rater-model}. Hence the proof is
immediate from \eqref{eq:basic-prob-Xk=c}.

Now recall the discussion at the end of Section~\ref{sec:model} and assume for a moment that $\beta$
would depend on $\gamma(k)$, so the original model would have the distribution
$$\Prob(X_k = c) = \beta_{\gamma(k)} \delta_{c,\gamma(k)} + (1-\beta_{\gamma(k)})p_{c}$$
\ie the mixture coefficient $\beta_{\gamma(k)}$ depends upon the support of the atomic measure.
Transforming the classes as in the preceding proposition, instead of
\eqref{eq:prop:category-mapping} we would arrive at the equation
$$\Prob(X_k' = c') = \beta_{\gamma(k)} \delta_{c',\gamma'(k)} + (1-\beta_{\gamma(k)})p_{c'}'$$
so $\beta$ no longer depends upon the supporting element $\gamma'(k)$ of the atomic measure alone.

Proposition~\ref{prop:category-mapping} provides a necessary condition for the validity of the
model: if one observes in an experiment that $\beta$ significantly changes when recomputed after
combining categories, the assumptions of the coder model are not met.  The numerical simulations in
the following section may give some indication which level of $\beta$-change could be considered as
significant.

Now we state and prove the main result on the identification of $\beta$ in the general case.

\begin{theorem}
  \label{thm:general-case} Let $(\beta,\gamma, \vec{p})$ be a coder model with coder processes $X_{i,k}$ for $i\in \N_R$,
  $k~\in~\N_N$. Moreover let $C^* = \{c\in C\,:\, e_{2,c} \not= e_{1,c}^2\}$. If $\tau_c < 1$ for
  all $c\in C$, then the following holds:

  \begin{enumerate}
  \item\label{thm:general-case-beta=0} $C^* = \emptyset$ if and only if $\beta = 0$.
  \item\label{thm:general-case-cstar-non-singleton} If $C^* \not= \emptyset$ then $\#C^* \geq 2$.    
  \item\label{thm:general-case-m-eq-2} If $\#C^* = 2$ then $\beta = \sqrt{ 4a+b^2}$, where 
    $$a := e_{2,c} - e_{1,c}^2 \text{ and } b := \frac{e_{3,c} - e_{1,c}^3}{e_{2,c}-e_{1,c}^2} - 3 e_{1,c}$$
    for some $c\in C^*$.          
  \item\label{thm:general-case-m-ge-3-e3} If $\#C^* \geq 3$ then
    \begin{equation}
          \beta = \frac{1}{\#C^*-2}\left( \sum_{c\in C^*} \frac{e_{3,c} - e_{1,c}^3}{e_{2,c}-e_{1,c}^2} + 3 \sum_{c\in C\setminus C^*} e_{1,c} - 3\right),\label{eq:thm:m-ge-3-e3}        
        \end{equation}
  \item\label{thm:general-case-m-ge-3-tau} Let $C^* = \{ c_1,\ldots{},c_{m^*}\}$ and assume $m^*\geq
    3$. For $i,j\in\N_{m^*}$ let $$\rho_{i,j} := \frac{e_{2,c_i,c_j}-e_{1,c_i} e_{1,c_j}}{e_{2,c_i,c_i}-e_{1,c_i}^2}.$$\\ 
    Then $\lambda\in\R^{m^*}$ is a solution of 
    \begin{align}
        0 & = \lambda_i \rho_{i,j} - \lambda_k \rho_{k,j} \text{ for all } i,j,k\in\N_{m^*} \text{
          with } i \not= j \not= k \label{eq:thm:lambda-rho-condition}\\
        {m^*}-1  & = \sum_{i=1}^{m^*}\lambda_i 
    \end{align}
     if and only if $\lambda_i = 1 - \tau_i$. Moreover, for the solution $\lambda_i$ the following holds
    \begin{equation}
     \beta = \sqrt{\frac{\sum_{c\in C} (e_{2,c}-e_{1,c}^2)}{1 - \sum_{j=1}^{m^*} (1-\lambda_j)^2}} \label{eq:thm:m-ge-3-tau}
   \end{equation}                   
  \end{enumerate}
\end{theorem}

\begin{Proof}
   Part \ref{thm:general-case-beta=0} is just a restatement of Proposition~\ref{prop:case-tau-known}\ref{prop:case-tau-known:item-beta=0}.

   By \ref{thm:general-case-beta=0} $C^* \not= \emptyset$ implies $\beta \not=0$ and by
   \eqref{eq:proof-uniform-category-case-joint-agreement-excess} $\tau_{c_0} > 0$ for some $c_0 \in
   C^*$. Since $\tau_{c_0} < 1$ and $\sum_{c\in C} \tau_c = 1$ there is $c_1 \in C$, $c_1 \not= c_0$
   with $\tau_{c_1} > 0$ and again by \eqref{eq:proof-uniform-category-case-joint-agreement-excess}
   $e_{2,c_1}-e_{1,c_1}^2 > 0$, so $c_1 \in C^*$ proving \ref{thm:general-case-cstar-non-singleton}.

  To prove part \ref{thm:general-case-m-eq-2} write $C^* = \{\tau_0, \tau_1\}$.  From
  \eqref{eq:def-e_3c} we see that for every $c\in C$
  \begin{align} 
    e_{3,c}-e_{1,c}^3 & = \beta^2\tau_c (1-\tau_c) \left( \beta (1+\tau_c) + 3 (1-\beta)p_c \right) \nonumber \\
                   & = (e_{2,c}-e_{1,c}^2) \left( \beta (1+\tau_c) + 3 (1-\beta)p_c \right) \label{thm:proof:e3delta}
  \end{align}
  using \eqref{eq:proof-uniform-category-case-joint-agreement-excess} above in the last step.  From
  \eqref{eq:proof-uniform-category-case-joint-agreement-excess} and \eqref{thm:proof:e3delta} we
  obtain $a = \beta^2 \tau_c (1-\tau_c) \geq 0$ and $b = \beta (1-2\tau_c)$, where $\tau_c=\tau_0$
  or $\tau_c = \tau_1$. Since $\tau_0 + \tau_1 = 1$ we see that $a$ is independent of $c$ that $b$
  is uniquely defined up to its sign. So $\sqrt{ 4a+b^2}$ is well defined and independent of the
  choice of $c$ in the definition of $a$ and $b$. Now
  $$ (1-2\tau_c)^2 a = (1-2\tau_c)^2\beta^2 \tau_c (1-\tau_c) = b^2 \tau_c (1-\tau_c)$$
  and thus
  \begin{equation}
        (4a+b^2)(\tau_c - \frac{1}{2})^2 =  \frac{b^2}{4}.\label{eq:thm:proof:tau0-eqn}  
  \end{equation}    
  Hence $4a+b^2 = 0$ implies $b^2=0$ and so $a=0$. Now $b=0$ if and only if $\beta=0$ or $\tau_c =
  \frac{1}{2}$ and $a=0$ if and only if $\beta=0$ or $\tau_c\in\{0,1\}$, which shows that $4a+b^2=0$
  if and only if $\beta=0$.
  
  On the other hand, if $4a+b^2 \not= 0$ we may solve \eqref{eq:thm:proof:tau0-eqn} for $\tau_c$
  and obtain 
  $$\tau_c = \frac{1}{2}\left(1 \pm \frac{b}{\sqrt{4a+b^2}}\right)$$
  and using the definition of $b$ (and that $\beta>0$) \ref{thm:general-case-m-eq-2} is proved.

  Now we prove \ref{thm:general-case-m-ge-3-e3}. From \eqref{thm:proof:e3delta} we obtain for each $c\in C^*$
  $$\frac{e_{3,c} - e_{1,c}^3}{e_{2,c}-e_{1,c}^2} =  \beta (1+\tau_c) + 3 (1-\beta)p_c. $$
  Now, since $C^* \not= \emptyset$ by part~\ref{thm:general-case-beta=0} $\beta \not= 0$. Hence for
  all $c\in C \setminus C^*$ we get $\tau_c = 0$ by
  Proposition~\ref{prop:case-tau-known}\ref{prop:case-tau-known:item-tau-or-beta=0}.  Thus
  $$\sum_{c\in C^*} \tau_c = \sum_{c\in C} \tau_c = 1$$ and that $e_{1,c} = (1-\beta)p_c$ for $c\in
  C\setminus C^*$.  This shows
  \begin{align*}
    \sum_{c\in C^*} \frac{e_{3,c} - e_{1,c}^3}{e_{2,c}-e_{1,c}^2} 
        &= \beta \sum_{c\in C^*} (1+\tau_c) + 3 (1-\beta) \sum_{c\in C^*} p_c \\
        &= \beta(\#C^*+1) + 3 (1-\beta) \left(1 - \sum_{c\in C\setminus C^*} p_c \right) \\
        &= \beta(\#C^*-2) + 3  - 3 \sum_{c\in C\setminus C^*} e_{1,c} 
  \end{align*}
  Since $\#C^* \not= 2$ we may solve for $\beta$ which finishes the proof of \ref{thm:general-case-m-ge-3-e3}.  

  Next we prove \ref{thm:general-case-m-ge-3-tau}.  As in the proof of
  part~\ref{thm:general-case-m-ge-3-e3} $\beta\not=0$ and $\tau_c=0$ if and only if $c\in C\setminus C^*$.
  
  Now combine \eqref{eq:proof-uniform-category-case-joint-agreement-excess}, \eqref{eq:def-e_1c},
  and \eqref{eq:def-e_2c1c2} to see that
  \begin{equation}                      
    \rho_{i,j} = \frac{\delta_{i,j}\tau_i-\tau_i \tau_j}{\tau_i (1-\tau_i)}\label{eq:thm:proof-rho}
  \end{equation}
  so the proof of the ``if''-part is obvious.

  Now assume that some $\lambda\in\R^{m^*}$ solves \eqref{eq:thm:lambda-rho-condition} and that
  $\sum_{i=1}^{m^*}\lambda_i = {m^*}-1$.  Since $\tau_i<1$ for all $i\in \N_m$ and
  $\sum_{i=1}^m \tau_i = \sum_{i=1}^{m^*} \tau_i = 1$ there are $i,k\in\N_{m^*}$, $i\not=k$ such
  that $\tau_i \not= 0 \not= \tau_k$.  So for all $j\in\N_{m^*}\setminus \{i,k\}$
  $$-\frac{\lambda_j}{1-\tau_j} = \frac{\lambda_j\rho_{j,i}}{\tau_i} = \frac{\lambda_k\rho_{k,i}}{\tau_i} = -\frac{\lambda_k}{1-\tau_k}$$
  and 
  $$-\frac{\lambda_j}{1-\tau_j} = \frac{\lambda_j\rho_{j,k}}{\tau_k} = \frac{\lambda_i\rho_{i,k}}{\tau_k} = -\frac{\lambda_i}{1-\tau_i}$$
  Since ${m^*}\geq 3$ the set $\N_{m^*}\setminus \{i,k\}$ is not empty and thus $\theta :=
  \frac{\lambda_i}{1-\tau_i} = \frac{\lambda_j}{1-\tau_j}$ holds for all $j\in\N_{m^*}$.  This shows
  that $\lambda_j = (1-\tau_j)\theta$ for all $j\in\N_{m^*}$.  Now 
  $${m^*}-1 = \sum_{i=1}^{m^*}\lambda_i = \theta ({m^*}-1)$$
  implies $\theta=1$, so $\lambda_i = 1 - \tau_i$.  Finally, since
  \begin{align*}
    \sum_{c\in C^*} (e_{2,c}-e_{1,c}^2) &= \beta^2 \sum_{c\in C^*} \tau_c (1-\tau_c) = \beta^2 (1- \sum_{c\in C^*} \tau_c^2)\\
                                  &= \beta^2 (1- \sum_{i=1}^{m^*} (1-\lambda_i)^2)
  \end{align*}
  and using that since $m^* >0$ the left hand side is positive so we may solve for $\beta$ and obtain \eqref{eq:thm:m-ge-3-tau}.
\end{Proof}

Using that $\tau_j \not= 0$ for $j\in\N_{m^*}$ we see from \eqref{eq:thm:proof-rho} that $\rho_{i,j}
\not= 0$ for $i,j\in\N_{m^*}$, so \eqref{eq:thm:lambda-rho-condition} can easily be solved by
forward substitution.  Experience shows, however, that computing $\beta$ according to
part~\ref{thm:general-case-m-ge-3-tau} is numerical unstable.  Its virtue lies in the fact that it shows
that $\beta$ is uniquely determined by double coincidence expectations $e_{2,i,j}$, which could be
estimated from the ratings of two coders, but only if $m \geq m^* \geq 3$, \ie if there are at least
three categories.

Parts~\ref{thm:general-case-m-ge-3-e3} and \ref{thm:general-case-m-eq-2} use the triple coincidence
expectations, which require the ratings of at least three raters but are applicable for all $m\geq 2$.

This raises the question of whether $\beta$ is uniquely determined given double coincidence expectation
values even in the case $m=2$.  The next proposition shows that this is not the case.

\begin{proposition}\label{prop:two-cat-e2-not-sufficient}
  Let $m=2$ and assume that $\tau_c < 1$ for $c\in C$.  Let $(\beta,\gamma, \vec{p})$ be a coder model with
  expectation values $e_{1,c_1}$, $e_{2,c_1,c_2}$, $c_1,c_2 \in C$, $e_1 :=
  max(e_{1,c_1},e_{1,c_2})$ and
  \begin{equation}
    \label{eq:prop:two-cat-e2-not-sufficient:I}
    I :=
    \begin{cases}
      [0,1]\text{, if } e_1 = 1\\
      [0,2(1-e_1)] \cup [1-e_1 + \frac{\beta^2\tau_{c_1} (1-\tau_{c_1})}{1-e_1},1]\text{, if }e_1 < 1
    \end{cases}
  \end{equation}
  Then if
  \begin{equation}
  \beta' \in [2\beta\sqrt{\tau_{c_1}(1-\tau_{c_1})},e_1 + \frac{\beta^2\tau_{c_1}(1-\tau_{c_1})}{e_1}] \cap I\label{eq:prop:two-cat-e2-not-sufficient:region-of-indeterminacy}
\end{equation}
and $\beta'^2 = 4\beta^2\frac{\tau_{c_1} (1-\tau_{c_1})}{1-\frac{n^2}{N^2}}$ for some $n\in \N_N$
such that $n+N\in 2\N$ then there is
a coder model $(\beta', \gamma\,', p'_{\cdot})$, where $\gamma\,'\,:\,\N_N~\rightarrow~C$ which
yields the same expectation values $e_{1,c_1}$, $e_{2,c_1,c_2}$, $c_1,c_2 \in C$.
\end{proposition}

\begin{Proof}
  Write $C=\{c_1,c_2\}$.  First observe that $e_{1,c_2} = 1 - e_{1,c_1}$, $e_{2,c_1,c_2} = e_{1,c_1} - e_{2,c_1,c_1}$
  and thus 
  $$e_{2,c_2,c_2} = e_{1,c_2} - e_{2,c_2,c_1} = 1 - 2 e_{1,c_1} + e_{2,c_1,c_1}.$$
  So we only need to show $e_{1,c_1} = e'_{1,c_1}$ and $e_{2,c_1,c_1} = e'_{2,c_1,c_1}$ for the
  corresponding expectations $e'_{2,c_1,c_1}$, $e'_{1,c_1}$ of the model $(\beta',\gamma',
  p'_{\cdot})$.  

  Let $c_0\in C$ be such that $e_{c_0} = e_1$.  Since $e_{1,c_1} + e_{1,c_2} = 1$ we
  have $e_1 = e_{c_0} \geq \frac{1}{2}$ .  We also abbreviate $e_2~:=~e_{2,c_0,c_0}$,
  $e'_1~:=~e'_{c_0}$, and $e'_2~:=~e'_{2,c_0,c_0}$, $\tau=\tau_{c_0}$, $p = p_{c_0}$,

  Observe also that $\tau_c<1$ for all $c\in C$ implies $0 < \tau_c <1$ for all $c\in C$ and that
  all conditions in the statement of the proposition are invariant if $\tau_{c_1}$ is exchanged for
  $\tau$.  Thus if $\beta' = 0$ also $\beta=0$ and the statement of the proposition is
  trivially satisfied in this case.  So for the following we may assume that $\beta'>0$.  By
  \eqref{eq:proof-uniform-category-case-joint-agreement-excess} $e_2-e_1^2 = \beta^2 \tau (1-\tau)$
  and by assumption $\beta' \geq 2\beta\sqrt{\tau (1-\tau)}$, so
  \begin{equation}
    \tau' := \frac{1}{2} + \frac{1}{2}\sqrt{ 1 - \frac{4\beta^2\tau (1-\tau)}{\beta'^2}}\label{eq:proof:prop:two-cat-e2-not-sufficient:def-tau}
  \end{equation}
  is well defined and satisfies 
  $$\beta'^2 \tau' (1-\tau') = \beta^2\tau (1-\tau)$$
  Hence 
  $$e'_2 = \beta'^2 \tau' (1-\tau') + {e'_1}^2 = \beta^2\tau (1-\tau) + {e'_1}^2 = e_2 + ({e'}_{1}^{2} - e_1^2),$$
  so we only need to prove 
  \begin{equation}
    e_1 = e'_1 = \beta'\tau' + (1-\beta')p'\label{eq:p-prime-equation}   
  \end{equation}
  Case $\beta' = 1$:  In this case from the assumption we see that 
  $$1 = \beta' \leq e_1 + \frac{\beta\tau (1-\tau)}{e1} = e_1 + \frac{e_2-e_1^2}{e_1} = \frac{e_2}{e_1} \leq 1$$
  (using $e_2 \leq \sum_{c\in C} e_{2,c_0,c} = e_1$),  so $e_2 = e_1$.  Now from \eqref{eq:def-e_2c}
  and \eqref{eq:def-e_1c}
  \begin{equation}
    \label{eq:e2me1-for-beta-prime-unitiy}
    0 = e_2 - e_1 = - (1-\beta) \left( \beta\tau (1-p) + \beta p (1-\tau) + (1-\beta)p (1-p) \right)
  \end{equation}
  Since every summand in the second factor of (\ref{eq:e2me1-for-beta-prime-unitiy}) is non-negative
  and $0 < \tau < 1$ this implies that either $\beta=0$ and $p\in \{0,1\}$ or $\beta=1$.

  First assume $\beta=0$. Since by assumption $0 < \frac{1}{2} \leq e_1 = \beta \tau + (1-\beta) p$
  only $p=1$ is possible. Now from \eqref{eq:proof:prop:two-cat-e2-not-sufficient:def-tau} we obtain
  $\tau' = 1$ and from $p=1$
  $$e_1 = \beta \tau + (1-\beta) p = p = 1 = \beta'\tau' = e'_1$$
  
  On the other hand, if $\beta=1$ we get $\tau = e_1 \geq \frac{1}{2}$ and so
  $${e'}_1 = \tau' = \frac{1}{2} + \frac{1}{2} \sqrt{1 - 4\tau (1-\tau)} = \tau = e_1.$$  
  This concludes the case $\beta' = 1$.

  If $\beta' < 1$ we may solve 
     $$e_1 = \beta'\tau' + (1-\beta')p'$$
  for $p'$ and it remains to show that $0\leq p' \leq 1$.

  From the assumption
  $$e_1 \beta' \leq e_1^2 + \beta^2\tau (1-\tau)$$
  and after reordering and completing the square we find that
  $$\sqrt{\beta'^2 - 4\beta\tau (1-\tau)} \leq | 2e_1 - \beta'| = 2e_1 - \beta'$$
  where the last equality follows from $\beta' < 1 \leq 2 e_1$.  This shows that
  $$\beta'\tau' = \frac{1}{2} \beta' + \frac{1}{2} \sqrt{\beta'^2 - 4\beta\tau (1-\tau)} \leq e1$$
  and thus $p' \geq 0$. 

  To prove $p' \leq 1$ first assume $e_1=1$.  Then 
  $$1-e_1 = \beta (1-\tau_{c_0}) + (1-\beta) (1-p_{c_0})$$
  and since $\tau_{c_0} < 1$ we conclude that $\beta=0$ and $p_{c_0}=1$. This implies $\tau'=1$ and
  so
  $$p' = \frac{e_1-\beta'\tau'}{1-\beta'} = \frac{1-\beta'}{1-\beta'} = 1.$$
  
  If $e_1 < 1$, by assumption, $\beta' \leq 2 (1-e_1)$ or $\frac{\beta^2\tau (1-\tau)}{1-e_1} +1-e_1
  \leq \beta'$ so again by reordering and completion of the square one sees that
  $$\beta'-2 (1-e_1) \leq 0 \text{ or } |\beta'-2 (1-e_1)| \leq \sqrt{\beta'^2 - 4\beta\tau (1-\tau)}$$
  \ie $\beta'-2 (1-e_1) \leq \sqrt{\beta'^2 - 4\beta\tau (1-\tau)}$ and thus
  $$e_1 - \beta'\tau' = e_1 - \frac{1}{2} \beta' - \frac{1}{2}\sqrt{\beta'^2 - 4\beta\tau (1-\tau)} \leq 1 - \beta'$$
  proving $p'\leq 1$.
  
  Finally, let $\beta'^2 = 4\beta^2\frac{\tau (1-\tau)}{1-\frac{n^2}{N^2}}$ for some $n\in
  \N_N$ such that $n+N\in 2\N$. Then by \eqref{eq:proof:prop:two-cat-e2-not-sufficient:def-tau}
  $$\tau' = \frac{1}{2} + \frac{1}{2} \sqrt{1 - 1 + \frac{n^2}{N^2}} = \frac{n+N}{2N}$$
  So $\tau' N$ is a natural number and \eg defining $\gamma\,'(k) = c_1$ for $k\leq \frac{n+N}{2}$ and
  $\gamma\,'(k) = c_2$ otherwise, completes the proof.
\end{Proof}

\begin{figure}
   \centering
   \includegraphics[width=0.55\textwidth,keepaspectratio=true]{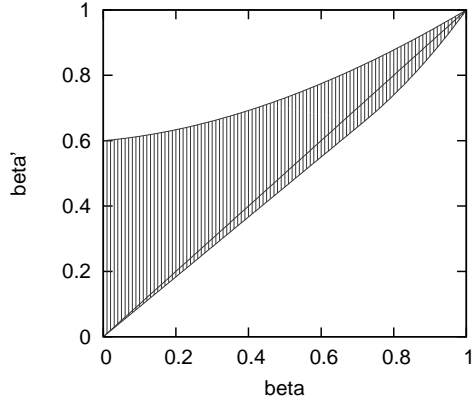}
   \caption{Example of the $\beta'$-region (shaded) according to
     \eqref{eq:prop:two-cat-e2-not-sufficient:region-of-indeterminacy} of
     Proposition~\ref{prop:two-cat-e2-not-sufficient}. The straight line inside the region indicates
     the value of $\beta$, \ie the diagonal.  Parameters are $\tau=(0.7,0.3)$, $p=(0.6,0.4)$}
   \label{fig:prop5-example}
\end{figure}

If in the preceding proposition $N$ is large enough several points of the set
 $$\{4\beta^2\frac{\tau (1-\tau)}{1-\frac{n^2}{N^2}}\;:\; n\in \N_{\eta N}, n+N\in 2\N\}$$
 (where $0<\eta<1$) fall into the set 
$$[2\beta\sqrt{\tau(1-\tau)},e_1 + \frac{\beta^2\tau(1-\tau)}{e_1}] \cap I$$
(if it has inner points) so the reliability parameter can not be determined uniquely.
Figure~\ref{fig:prop5-example} shows the $\beta'$-range given by
\eqref{eq:prop:two-cat-e2-not-sufficient:region-of-indeterminacy} for some random example.

So in the two-category case we need an estimate of $e_{3,c}$ in order to apply part
\ref{thm:general-case-m-eq-2} of Theorem~\ref{thm:general-case}, \ie we need at least three coders
to determine $\beta$ in this case.

As a consequence for the popular two-coder/two-category examples $\beta$ (more precisely the triple
$(\beta,\tau,\vec{p})$) is not uniquely determined.  In order to determine $\beta$ in such a situation we
thus either need to know $\vec{p}$ and use Proposition~\ref{proposition:special-cases} or
$\vec{\tau}$ and apply Proposition~\ref{prop:case-tau-known}.

\section{Numerical Simulations}
\label{sec:numerical-simulation}

The formulae provided by Theorem~\ref{thm:general-case} involve expectation values of coder agreement
frequencies.  In experiments we typically do not know expectation values but rather observe
relative frequencies. Hence we will not obtain the correct values for $\beta$ using the formulae in
parts \ref{thm:general-case-m-eq-2} and \ref{thm:general-case-m-ge-3-e3} of
Theorem~\ref{thm:general-case}.  Actually, these formulae involve differences of expectation values
which are close to 0 for small values of $\beta$, so small statistical fluctuations might lead to
large deviations in $\beta$.  Thus in order to improve the accuracy we reformulate the problem as a
least square optimization problem for the expectation values $e_{1,c}$,$e_{2,c_1,c_2}$ and (if $\#C
> 2$) $e_{3,c}$, using the formulae for $\beta$ to obtain a start value (augmented by approximations
for $\tau$ and $p$ according to \eqref{eq:proof-uniform-category-case-joint-agreement-excess} and \eqref{eq:def-e_1}
respectively). So find $\beta$, $\tau$, $p$ such that 

\begin{eqnarray*}
  \label{eq:optimization-target}
  & & \sum_{c\in C}  \left(\beta\tau_c + (1-\beta)p_c -e_{1,c}\right)^2 \\
  & & + \sum_{c_1,c_2\in C} \left(\beta^2\tau_c + 2\beta(1-\beta)p_c\tau_c + (1-\beta)^2p_c^2 -e_{2,c_1,c_2}\right)^2 \\
  & & + \sum_{c\in C}  \left(\beta^3\tau_c + 3\beta^2(1-\beta)p_c\tau_c + 3\beta(1-\beta)^2p_c^2\tau_c +
    (1-\beta)^3p_c^3 -e_{3,c}\right)^2
\end{eqnarray*}

is minimized, subject to the natural constraints.  

As we already noted in the introduction the model based approach chosen here allows for simulation
runs to investigate the accuracy of this algorithm.  The remainder of this section is devoted to such
numerical experiments which show the accuracy with varying model parameters. We display
the results as inverse empirical distribution functions for a sample of 1000 randomly chosen
realizations of the coder model, so the abscissae contain the quantiles and the ordinates the
absolute errors (observe the ranges).  In the plots the values for the 50~\%, 80~\%, 90~\%,
95~\%, 98~\%, and 100~\% quantiles are highlighted.  For every plot we also indicate the other
parameters in the caption.  The meaning of the parameters is that of the coder model defined in
Section~\ref{sec:model}.

\begin{figure}[]
   \centering
   \includegraphics[width=0.55\textwidth,keepaspectratio=true]{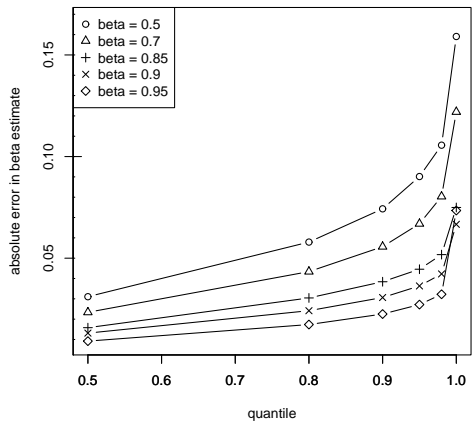}
   \caption{Estimation errors as a function of the true $\beta$ value. 
   Fixed parameters: $N=100$, $m=3$, $R=5$, $\tau =(0.3,0.6,0.1)$, $p=(0.33,0.33,0.34)$ }
   \label{fig:fig1-error-vs-beta}
\end{figure}

The accuracy in $\beta$ estimation depends on the actual value of $\beta$.  As Figure
\ref{fig:fig1-error-vs-beta} shows, the error decreases with increasing true value of $\beta$. The
98\% quantile accuracy ranges from $0.032$ at $\beta_{true}=0.95$ to $0.105$ at $\beta_{true}=0.5$.

According to the coder model a value of $\beta=0.5$ means that only for half of the items the raters
could determine the categories with certainty.  Note also that if the assumptions of
Proposition~\ref{proposition:beta-estimate} are satisfied the $S$-value would be as low as 0.25 in
this case.  Hence the really interesting range for $\beta$ is above 0.5 where the accuracy is
higher. 

\begin{figure}[]
   \centering
   \includegraphics[width=0.55\textwidth,keepaspectratio=true]{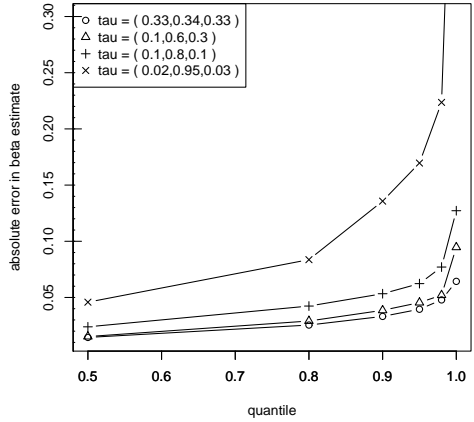}
   \caption{ Error of $\beta$ estimate for different true class frequencies.
   Fixed parameters: $N=100$, $m=3$, $R=5$, $\beta = 0.85$, $p=(0.33,0.33,0.34)$}
   \label{fig:fig2-error-vs-tau}
\end{figure}

As has been noted before, the definition of $\beta$ does not exhibit the ``problem of prevalence'',
\ie its value does not decrease when $\max(\tau)$ approaches $1$.  Though the value of $\beta$ is
not affected it does affect the accuracy as Figure \ref{fig:fig2-error-vs-tau} shows.  The 98\%
quantile accuracy ranges from $0.032$ for $\max(\tau)=\frac{1}{3}$ (the least value of $\max(\tau)$
in this setting) to $0.077$ for $\max(\tau)=0.90$ and $0.22$ for $\max(\tau)=0.95$.  In this latter
case there are only five of the items not belonging to the prevalent category.  Hence statistical
fluctuations may blur the distinction of this case from the case $\max(\tau)=1$ where $\beta$ can no
longer be determined according to Proposition~\ref{proposition:single-category-case}.  So this
decrease in accuracy is expected.

\begin{figure}[]
  \centering
   \includegraphics[width=0.55\textwidth,keepaspectratio=true]{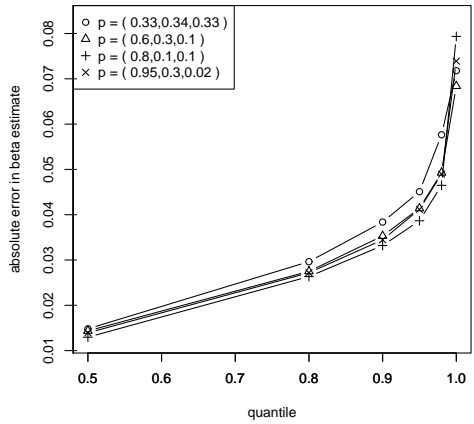}
   \caption{Error of $\beta$ estimate for different a priori distributions.
   Fixed parameters: $N=100, m=3, R=5, \beta = 0.85, \tau=(0.3,0.6,0.1)$}
   \label{fig:fig3-error-vs-p}
\end{figure}

Contrary to the rather strong impact of $\tau$ on the accuracy, the \textit{a priori} distribution $p$
does no significantly influence the accuracy as the following Figure \ref{fig:fig3-error-vs-p}
shows.  Here the 98\% quantile errors range from $0.049$ to $0.058$ which may be fully
attributed to statistical fluctuations.

\begin{figure}[]
   \centering
   \includegraphics[width=0.55\textwidth,keepaspectratio=true]{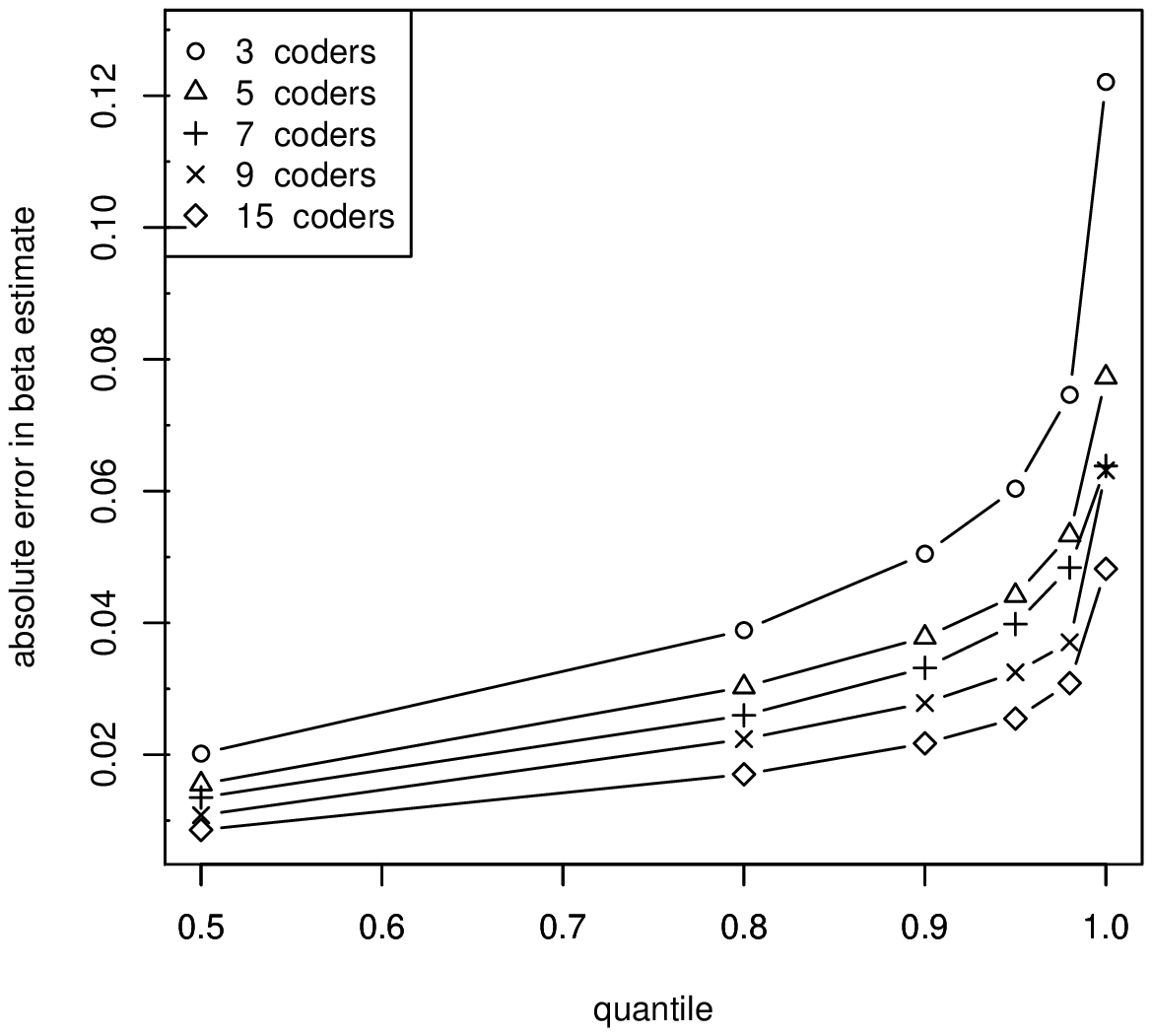}
   \caption{Error of $\beta$ estimate for different number of coders.
   Fixed parameters: $N=100, m=3, \beta = 0.85, \tau=(0.3,0.6,0.1), p = (0.33,0.33,0.34)$}
   \label{fig:fig4-error-vs-number-of-coders}
\end{figure}

\begin{figure}[]
   \centering
   \includegraphics[width=0.55\textwidth,keepaspectratio=true]{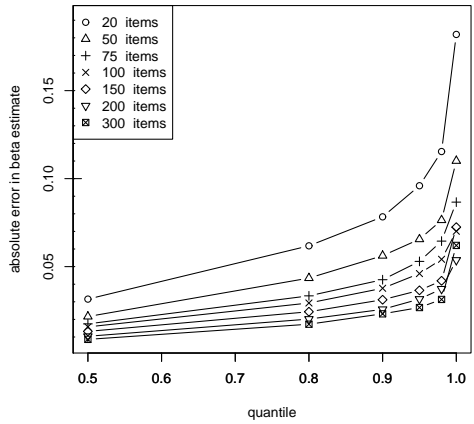}
   \caption{Error of $\beta$ estimate for different number of items.
   Fixed parameters: $m=3, R=5, \beta = 0.85, \tau=(0.3,0.6,0.1), p = (0.33,0.33,0.34)$}
   \label{fig:fig5-error-vs-number-of-items}
 \end{figure}

Finally, since the errors originate from deviations of relative frequencies from the expectation
values, the accuracy depends of course moderately on both the number of coders and the number of
items.  Figure \ref{fig:fig4-error-vs-number-of-coders} shows the influence of the number of coders
on the accuracy.  At the 98\% quantile level the errors range from $0.03$ (15 coders) to $0.07$ (3
coders).  Actually as few as five coders suffice to obtain a reasonably low error of $0.053$.

The impact of the number of items can be seen from Figure \ref{fig:fig5-error-vs-number-of-items}:
With as few as $20$ items one cannot expect more than a rough estimate of beta (error $0.115$ at
98\% quantile) with a reasonably low error of $0.054$ when coding 100 items.

\bibliographystyle{plain}
\bibliography{ref} 
\end{document}